\documentclass{PoS}

\usepackage{amssymb}
\usepackage{fontenc}
\usepackage{times}
\usepackage{mathptmx}
\usepackage{graphicx}
\usepackage{cite}

\usepackage{amsmath}

\usepackage[english]{babel}
\usepackage[latin1]{inputenc}

\usepackage{epsfig}
\usepackage{subfigure}

\def\qbar{{\overline{q}}}

\def\qbold{{\mathbf{q}}}

\def\Nbar{{\bar{N}}}

\def\alfabar{{\bar{\alpha}}}
\def\betabar{{\bar{\beta}}}

\def\Dtilde{\tilde{D}}

\newcommand{\bea}{\begin{eqnarray}}
\newcommand{\eea}{\end{eqnarray}}
\newcommand{\be}{\begin{equation}}
\newcommand{\ee}{\end{equation}}

\title{The double gluon distribution from the single gluon distribution}

\ShortTitle{double Gluon Distribution}

\author{Krzysztof Golec Biernat\\
                Institute of nuclear Physics, Polish Academy of Sciences\\
Faculty of Mathematics and Natural Sciences, University of Rzesz\'ow}

\author{Emilia Lewandowska \\
Institute of Nuclear Physics, Polish Academy of Sciences}

\author{\speaker{Mirko Serino}
\\
        Institute of nuclear Physics, Polish Academy of Sciences}

\author{Zachary Snyder\\
Penn State University}

\author{Anna M. Sta\'sto \\
Penn State University, University Park, PA 16802, United States\\
Institute of Nuclear Physics Polish Academy of Sciences}
        
\abstract{
Using momentum sum rule for evolution equations for Double Parton Distribution Functions (DPDFs)
in the leading logarithmic approximation, we find that the double gluon distribution
function can be uniquely constrained via the single gluon distribution function.
We also study numerically its evolution with a hard scale and show that an approximately factorized ansatz 
into the product of two single gluon distributions performs quite well at small values of $x$ 
but is always violated for larger values, as expected.
}

\FullConference{XXIV International Workshop on Deep-Inelastic Scattering and Related Subjects\\
11-15 April, 2016 \\
DESY Hamburg, Germany}

\begin{document}
  
\section{Evolution equations and sum rules}

In the collinear leading logarithmic approximation double parton distribution functions (DPDFs) obey QCD evolution equations 
\cite{Kirschner:1979im,Shelest:1982dg,Zinovev:1982be,Snigirev:2003cq,Korotkikh:2004bz,Gaunt:2011xd} which are
similar to the Dokshitzer-Gribov-Lipatov-Altarelli-Parisi (DGLAP) equations for single parton distribution functions (PDFs). 
The evolution equations for DPDFs conserve new sum rules which relate
the double and single parton distributions and are conserved by the evolution. 

We consider the DPDFs with equal hard scales, $Q_1=Q_2\equiv Q$ and null relative momentum $\qbold=0$: 
\be
D_{f_1f_2}(x_1,x_2,Q)\,\equiv\,D_{f_1f_2}(x_1,x_2,Q,Q,\qbold=0)\; ,
\label{eq:dpdfdef}
\ee
where $x_{1,2}\in [0,1]$ are parton momentum fractions obeying the condition $x_1+x_2\le1$  
and $f_{1,2}$ denote parton species. 
With this simplifying assumptions, 
the evolution equations in the  leading logarithmic approximation read 
\begin{eqnarray}
&&
\frac{\partial}{\partial {\ln Q^2}}\, D_{f_1f_2}(x_1,x_2,Q) =
\frac{\alpha_s(Q)}{2\pi}\sum_{f'} \Bigg\{\int^{1-x_2}_{x_1} \frac{du}{u} \,{\cal{P}}_{f_1f'}\!\left(\frac{x_1}{u}\right) D_{f' f_2}(u,x_2,Q)
\nonumber \\
&& \hspace{-10mm}
+\int_{x_2}^{1-x_1} \frac{du}{u}\,{\cal{P}}_{f_2f'}\!\left(\frac{x_2}{u}\right)D_{f_1f'}(x_1,u,Q)
+ \frac{1}{x_1+x_2}\,{P}^R_{f'\to f_1f_2}\!\left(\frac{x_1}{x_1+x_2}\right) D_{f'}(x_1+x_2,Q)\Bigg\}.
\label{eq:twopdfeq}
\end{eqnarray}
where the functions  $\cal{P}$ on the r.h.s. are  the leading order Altarelli-Parisi splitting functions (with virtual corrections for ${\cal{P}}_{ff}$ included). 
The third term on the r.h.s corresponds to the splitting of one parton into two daughter partons, 
described by the Altarelli-Parisi splitting function for real emission, $P^R_{f'\to f_1f_2}$. 
As  eq.~\eqref{eq:twopdfeq} contains the single PDFs, $D_{f'}$, 
it has to be solved together with the ordinary DGLAP equations, see e.g. Ref.~\cite{Gaunt:2011xd} for more details.
The splitting terms in the evolution equations are crucial for the conservation of sum rules which are discussed below.

The sum rules which are conserved by the  evolution equations (\ref{eq:twopdfeq})  are the momentum and valence quark
number sum rules \cite{Gaunt:2009re},
\begin{eqnarray}
&&
\sum_{f_1}\int_{0}^{1-x_2}dx_1\,x_1D_{f_1f_2}(x_1,x_2) =
(1-x_2)D_{f_2}(x_2) \; ,
\label{eq:momrule1}
\\
&&
\int_0^{1-x_2}dx_1\!\left\{D_{qf_2}(x_1,x_2)-D_{\qbar f_2}(x_1,x_2)\right\}  =
(N_q-\delta_{f_2q}+\delta_{f_2\qbar}) D_{f_2}(x_2)\,,
\label{eq:valrule1}
\end{eqnarray}
where $q=u,d,s$ and  $N_u=2, N_d=1,N_s=0$ are the valence quark number for each of the quark flavors. 
The same relations hold true with respect to the second parton, as the DPDFs are  parton exchange symmetric,
\be
\label{eq:pes}
D_{f_1f_2}(x_1,x_2)=D_{f_2f_1}(x_2,x_1)\, .
\ee

We see that the above sum rules relate the double and single parton distribution
functions, which reflects the common origin of those distributions, namely the  expansion of the nucleon state in Fock
light-cone components \cite{Gaunt:2009re}. 
In addition,  the sum rules for the single parton distributions are also satisfied - 
the momentum and quark valence sum rules for  $q=u,d,s$
\begin{eqnarray}
\sum_f\int_0^1dx\,xD_f(x) &=& 1
\nonumber \\
\int_0^1dx\,\left\{D_q(x)-D_{\qbar}(x)\right\} &=& N_q\,.
\end{eqnarray}

Let us introduce the Mellin transforms of single and double PDFs
\begin{eqnarray}
\Dtilde_f(n) 
&=& 
\int_0^1dx\,  x^{n-1}D_f(x)\, , 
\nonumber \\
\label{eq:singlemellin}
\Dtilde_{f_1f_2}(n_1,n_2) &=& 
\int_0^1dx_1\int_0^1dx_2\,  (x_1)^{n_1-1}(x_2)^{n_2-1} {D}_{f_1f_2}(x_1,x_2) \Theta(1-x_1-x_2).
\label{eq:doublemellin}
\end{eqnarray}
where $n$ and $n_{1,2}$  are complex numbers and we omit the scale $Q_0$ in the notation from now on.

The sum rules (\ref{eq:momrule1}) and (\ref{eq:valrule1}) can be written with the help of the Mellin moments 
after the  integration of both sides over $x_2$ with the factor $(x_2)^{n_2-1}$. 
Thus we find
\begin{eqnarray}
\label{eq:mom1}
\sum_{f_1}\,\Dtilde_{f_1f_2}(2,n_2) &=& \Dtilde_{f_2}(n_2) - \Dtilde_{f_2}(n_2+1)\, ,
\\\nonumber
\\
\Dtilde_{q f_2}(1,n_2) - \Dtilde_{\bar{q}f_2}(1,n_2) &=& (N_q-\delta_{f_2 q}+\delta_{f_2\qbar}) 
\Dtilde_{f_2}(n_2). 
\label{eq:val1}
\end{eqnarray}
Analogous relations are true for the  second parton.
These sum rules have to be satisfied simultaneously with 
the momentum and valence quark sum rules for the single parton distribution
\be
\label{eq:momsingle}
\sum_f \tilde{D}_f(2) = 1\, , \quad 
\tilde{D}_q(1)-\tilde{D}_\qbar(1)=N_q \, .
\ee
We want to construct initial conditions for DPDFs which fulfill the above sum rules since the PDFs on the r.h.s of 
Eqs.~(\ref{eq:momrule1})-(\ref{eq:valrule1}) are very well known from the global analysis fits. 
Thus, the PDFs constrain the DPDFs, significantly reducing the problem of uncertainty in the specification of initial conditions for DPDFs evolution.
For this purpose, we consider the LO single PDF parametrization from the MSTW fits \cite{Martin:2009iq}, 
since the evolution equations (\ref{eq:twopdfeq}) are given in  leading logarithmic approximation.

\section{Pure gluon case}

The single gluon distribution is specified in the LO MSTW parameterization at the scale $Q_0=1~{\rm GeV}$ and is given in the form 
\be\label{eq:mstwg}
D_g(x) = A_g\,x^{\delta_g-1} (1-x)^{\eta_g} (1 + \epsilon_g\,\sqrt{x} + \gamma_g\,x) \, ,
\ee

The parametrization  (\ref{eq:mstwg}) can be written in a general form which is more suitable for our purpose
\be\label{eq:newmstwg}
D_g(x) \; = \; \sum_{k=1}^L N^k_g\, x^{\alpha^k_g} \, (1-x)^{\beta_g^k}\,,
\ee
where $L=3$. In the Mellin space, the gluon distribution \eqref{eq:mstwg} can be written as
\begin{equation}
\tilde{D}_g(n) \; = \;  \sum_{k=1}^{L} N^k_g \, \frac{\Gamma(n+\alpha^k_g)\Gamma(\beta^k_g+1)}{\Gamma(n+\alpha^k_g+\beta^k_g+1)} \; .
\label{eq:dnsumk}
\end{equation}
where the expression on the r.h.s., $\Gamma(x)\Gamma(y)/\Gamma(x+y)\equiv B(x,y)$, is the  Euler Beta function. 
Thus the MSTW parametrization for the initial condition is in the form of the sum over the Beta functions with different sets of 
parameters which govern the small $x \rightarrow 0$ and large $x\rightarrow 1$ behavior. 

For the double parton distribution 
$D_{gg}$ the ansatz we take is the sum over the Dirichlet-type distributions of order $K=3$
\be\label{eq:gg1}
D_{gg}(x_1,x_2) \; = \; \sum_{k=1}^L \Nbar_{gg}^k\,(x_1x_2)^{\alfabar_g^k}\,(1-x_1-x_2)^{\betabar_g^k} \, ,
\ee
where $ \Nbar_{gg}^k, \alfabar_g^k$ and $\betabar_g^k$ are the parameters to be determined.
with $x_1,x_2 >0, x_1+x_2 \le 1$ and $x_3 = 1-x_1-x_2$.

The detail of the construction procedure can be found in \cite{Golec-Biernat:2015aza}. Here we give the final result,
i.e. the parameter-free double gluon distribution at the initial scale $Q_0=1~{\rm GeV}$,
\be\label{eq:initialgg}
D_{gg}(x_1,x_2)=\sum_{k=1}^3 \, N^k_g\frac{\Gamma(\beta^k_g+2)}{\Gamma(\alpha^k_g+2)\Gamma(\beta^k_g-\alpha^k_g)}\,
(x_1x_2)^{\alpha^k_g}\,(1-x_1-x_2)^{\beta^k_g-\alpha^k_g-1} \, .
\ee
%

\begin{figure*}[t]
\includegraphics[width = 10cm]{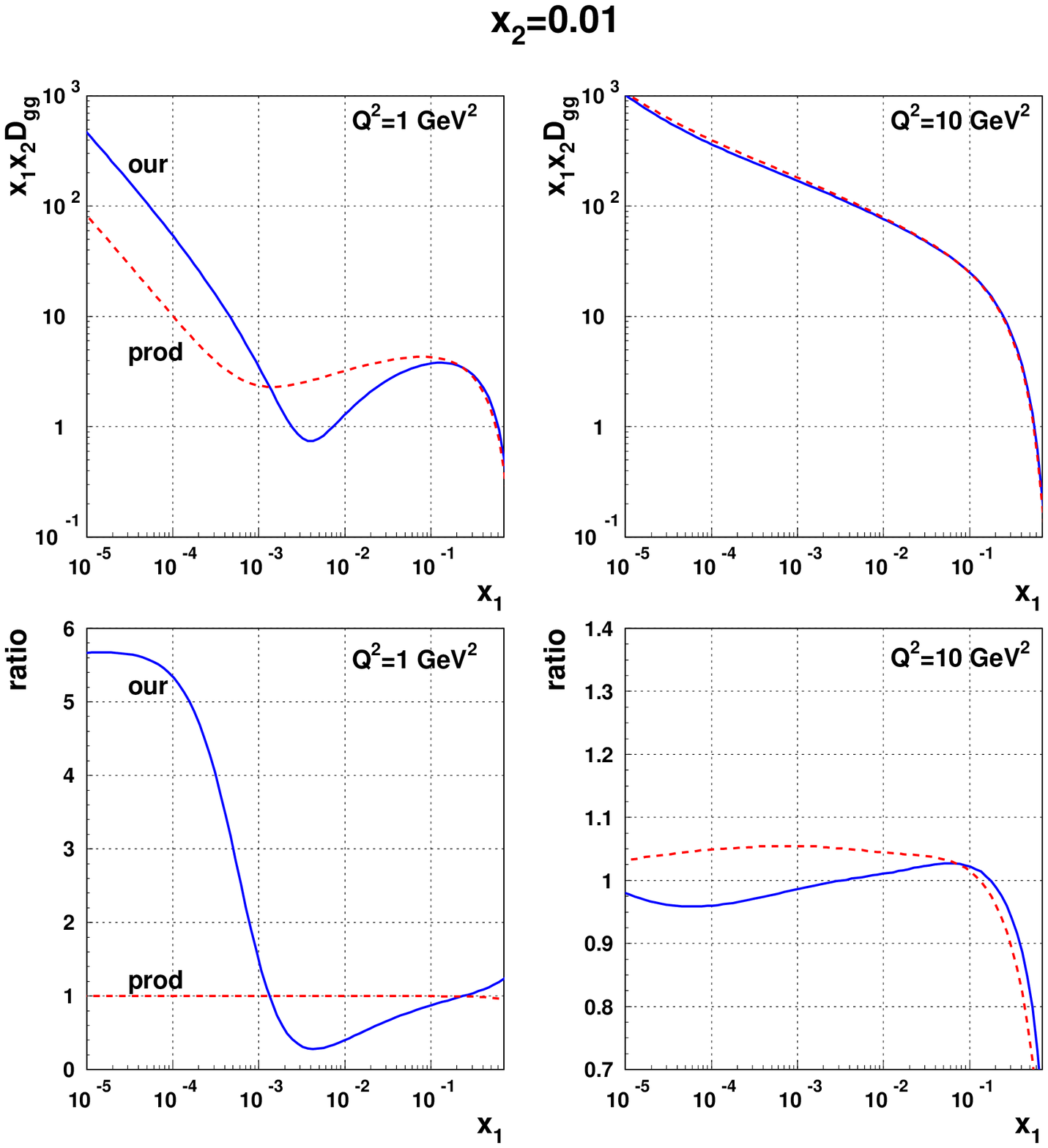}
\caption{The distribution $x_1x_2D_{gg}(x_1,x_2=10^{-2})$ at  $Q_0^2=1~{\rm GeV}^2$ (left upper panel) and 
$Q^2=10~{\rm GeV}^2$ (right upper panel) and the ratio \eqref{eq:ratio} (lower panels). The solid lines correspond to input (\ref{eq:initialgg}) (our)
while the dashed lines to input (\ref{eq:gaunt}) (prod).
}
\label{fig1}
\end{figure*}

\begin{figure*}[t]
\includegraphics[width = 10cm]{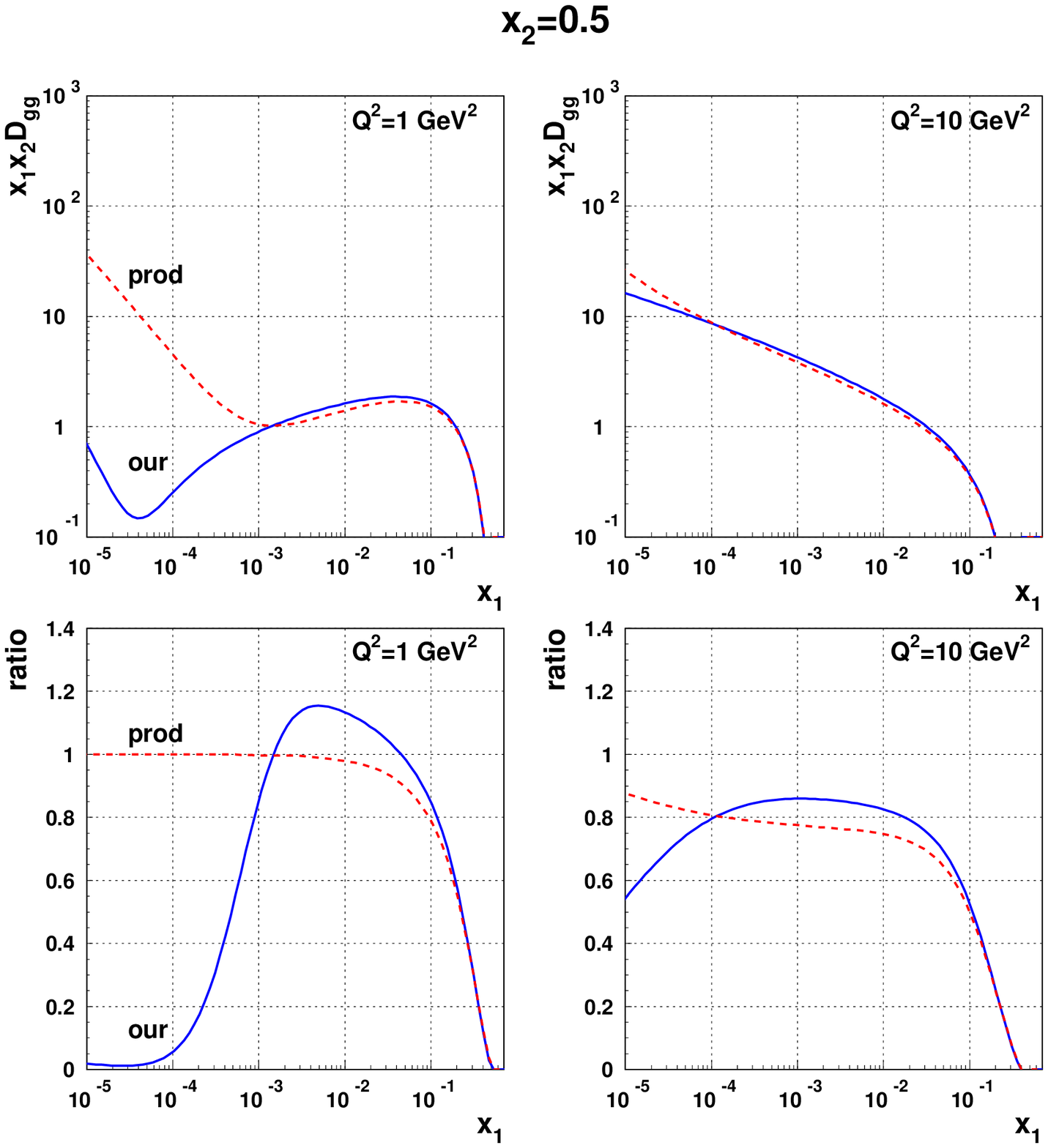}
\caption{The distribution $x_1x_2D_{gg}(x_1,x_2=0.5)$ at  $Q_0^2=1~{\rm GeV}^2$ (left upper panel) and 
$Q^2=10~{\rm GeV}^2$ (right upper panel) and the ratio \eqref{eq:ratio} (lower panels). The solid lines correspond to input (\ref{eq:initialgg}) (our)
while the dashed lines to input (\ref{eq:gaunt}) (prod).
}
\label{fig2}
\end{figure*}


The evolution equations (\ref{eq:twopdfeq})  reduced to the pure gluon case have the following form
\begin{eqnarray}\nonumber
\label{eq:twopdfeqgg}
&&
\frac{\partial}{\partial {\ln Q^2}}\, D_{gg}(x_1,x_2,Q) = \frac{\alpha_s(Q)}{2\pi}
\Bigg\{\int^{1-x_2}_{x_1} \frac{du}{u} \,{\cal{P}}_{gg}\!\left(\frac{x_1}{u}\right) D_{gg}(u,x_2,Q)
\nonumber \\
&& \hspace{-10mm}
+\int_{x_2}^{1-x_1} \frac{du}{u}\,{\cal{P}}_{gg}\!\left(\frac{x_2}{u}\right)D_{gg}(x_1,u,Q)
+ \frac{1}{x_1+x_2} \,{P}^R_{gg}\!\left(\frac{x_1}{x_1+x_2}\right) D_{g}(x_1+x_2,Q)\Bigg\}.
\end{eqnarray}
where $P^R_{gg}$ is the gluon-to-two gluon splitting function for real emission in the LO approximation. 
Strictly speaking, such an equation can be a reasonable approximation for small values of the momentum fractions,
where the gluons are known to dominate.
We solve numerically the above equation with the initial condition (\ref{eq:initialgg}).  We compare our results with 
those obtained from the usually assumed form of the initial conditions \cite{Gaunt:2011xd}, which satisfy  the momentum sum rule only approximately,
\be\label{eq:gaunt}
D_{gg}(x_1,x_2)=D_g(x_1)D_g(x_2)\rho(x_1,x_2) \, 
\quad \text{with}
\rho(x_1,x_2)=\frac{(1-x_1-x_2)^2}{(1-x_1)^2(1-x_2)^2} \, .
\ee 

The results are shown in Fig.~\ref{fig1} and Fig.~\ref{fig2}. We plot there the double gluon distribution $x_1x_2D_{gg}(x_1,x_2)$ as a function of $x_1$ 
for two values of the scale, initial $Q_0^2=1~{\rm GeV}^2$ and $Q^2=10~{\rm GeV}^2$ (upper panels), for two fixed values of $x_2$,
respectively big ($10^{-2}$) and large ($0.5$). 
The solid lines show the results obtained from our input
(\ref{eq:initialgg}) while the dashed lines correspond to the input (\ref{eq:gaunt}) with the gluon distribution (\ref{eq:mstwg}). In the lower panels we plot the ratio,
\be
\label{eq:ratio}
{\rm ratio}=\frac{D_{gg}(x_1,x_2)}{D_g(x_1)D_g(x_2)}\,,
\ee
which characterizes factorizability  of the double gluon distribution into a product of two single gluon distributions. 

For both values of $x_2$, the initial double gluon distributions differ significantly for small values of $x_1$, up to $10^{-1}$ for $x_2=10^{-2}$ and
up to $10^{-3}$ for $x_2=0.5$. However, the QCD evolution equation erases this difference already 
at the scale $Q^2=10~{\rm GeV}^2$, see the upper panels in both figures.
As we have observed, the initial distribution (\ref{eq:initialgg}) is not factorizable into a product of two single gluon distributions for any values of
$x_1$ and $x_2$. However, if both momentum fractions are small ($<0.01$,) $D_{gg}$ becomes factorizable 
with good accuracy after evolution to the shown value of $Q^2$, see the lower panels in both figures. 
A small  breaking of the factorization can be attributed to the non-homogeneous term in the evolution equation \eqref{eq:twopdfeqgg}. 
If one of the two momentum fractions is large, like the shown $x_2=0.5$, this is no longer the case and the factorization is significantly broken
for all values of $x_1$ independent of the values of the evolution scale. 
We check that for larger values of $Q^2$ that than shown here.
We have to remember, however, that the large $x$ domain has to be supplemented by quarks
before any possibly phenomenologically relevant claims can be made.

\section{Summary}
We showed how to obtain a parameter free double gluon distribution $D_{gg}$ from the known single gluon distribution $D_g$, 
given by the MSTW parameterization, in the pure gluon case, using minimal hypotheses.
The next step would be to extend this formalism to include the quarks and satisfy the momentum and valence quark sum rules simultaneously. 

\section*{Acknowledgments}
This work was supported by the Polish NCN Grants No.~DEC-2011/01/B/ST2/03915 and DEC-2013/10/E/ST2/00656,  
by the Department of Energy  Grant No. DE-SC-0002145, by the Center 
for Innovation and Transfer of Natural Sciences and Engineering Knowledge in Rzesz\'ow and by the Angelo Della Riccia foundation.

\end{document}